\documentclass[prl,twocolumn,showpacs,preprintnumbers,amsmath,amssymb,superscriptaddress]{revtex4}
\usepackage[english]{babel}

\usepackage{graphicx}
\usepackage{color}
\usepackage[colorlinks,bookmarks=false,citecolor=blue,linkcolor=red,urlcolor=blue]{hyperref}
\usepackage{times}

\def\ket#1{|\,#1\,\rangle}

\def\opone{\le\textbf{}\textbf{}avevmode\hbox{\small1\kern-3.8pt\normalsize1}}

\begin{document}

\author{Andreas~M.~L\"auchli}
\affiliation{Max-Planck-Institut f\"{u}r Physik komplexer Systeme, N\"{o}thnitzer Stra{\ss}e 38, D-01187
 Dresden, Germany}
\author{Emil J. Bergholtz}
\affiliation{Max-Planck-Institut f\"{u}r Physik komplexer Systeme, N\"{o}thnitzer Stra{\ss}e 38, D-01187
 Dresden, Germany}
\author{Juha Suorsa}
\affiliation{Department of Physics, University of Oslo, P.O. Box 1048 Blindern, 0316 Oslo, Norway}
\author{Masudul Haque}
\affiliation{Max-Planck-Institut f\"{u}r Physik komplexer Systeme, N\"{o}thnitzer Stra{\ss}e 38, D-01187
 Dresden, Germany}

\title{Disentangling Entanglement Spectra of Fractional Quantum Hall States on Torus Geometries}

\date{\today}

\begin{abstract} 

We analyze the entanglement spectrum of Laughlin states on the torus and show
that it is arranged in towers, each of which is generated by modes of two
spatially separated chiral edges. This structure is present for all torus
circumferences, which allows for a microscopic identification of the prominent
features of the spectrum by perturbing around the thin torus limit.
\end{abstract}

\maketitle

\paragraph*{Introduction ---}
The description of condensed matter phases using entanglement measures,
borrowed from the field of quantum information theory, has led to an explosive
growth of interdisciplinary work \cite{AmicoFazioOsterlohVedral_RMP08}.
Despite all this interest, there are few cases where entanglement concepts
provide physical information that is not obtainable through more conventional
quantities, such as correlation functions.  One 
example involves topologically ordered states,
for which  bipartite entanglement measures 
have been shown to be useful probes \cite{kitaev06, levin06, LiHaldane_PRL08}.
Fractional quantum Hall (FQH) states of two-dimensional electrons in a
magnetic field stand out as 
%the only 
experimentally realized topologically
ordered phases,
%These states 
and have recently received renewed intense attention partly due to
quantum computation proposals based on their topological properties
\cite{topol-quantum-computing}.  An intriguing feature of FQH states is that
their edges have gapless modes, described by chiral Luttinger liquids
\cite{Wen_chiraledges, edgeNumerics_XinWan_Jain}.  In this work we study the
interplay of two edges, through the study of \emph{entanglement spectra}.

We focus on bipartite entanglement between two parts ($A$ and $B$) of the
system. 
% in its ground state $\ket{\psi}$.  
The entanglement spectrum (ES),
$\{\xi_i\}$, is defined in terms of the Schmidt decomposition
\begin{equation*}
\ket{\psi}=\sum_i e^{-\xi_i/2}\ket{\psi_i^A}\otimes\ket{\psi_i^B}. 
\end{equation*}
Here $\ket{\psi}$ is the ground state, and the states
$\ket{\psi_i^A}$ ($\ket{\psi_i^B}$) form an orthonormal basis for the
subsystem $A$ ($B$).

%%%%%%%%%%%%%%%%%%%%%%%%%%
\begin{figure}[t]
\centerline{\includegraphics[width=0.7\linewidth]{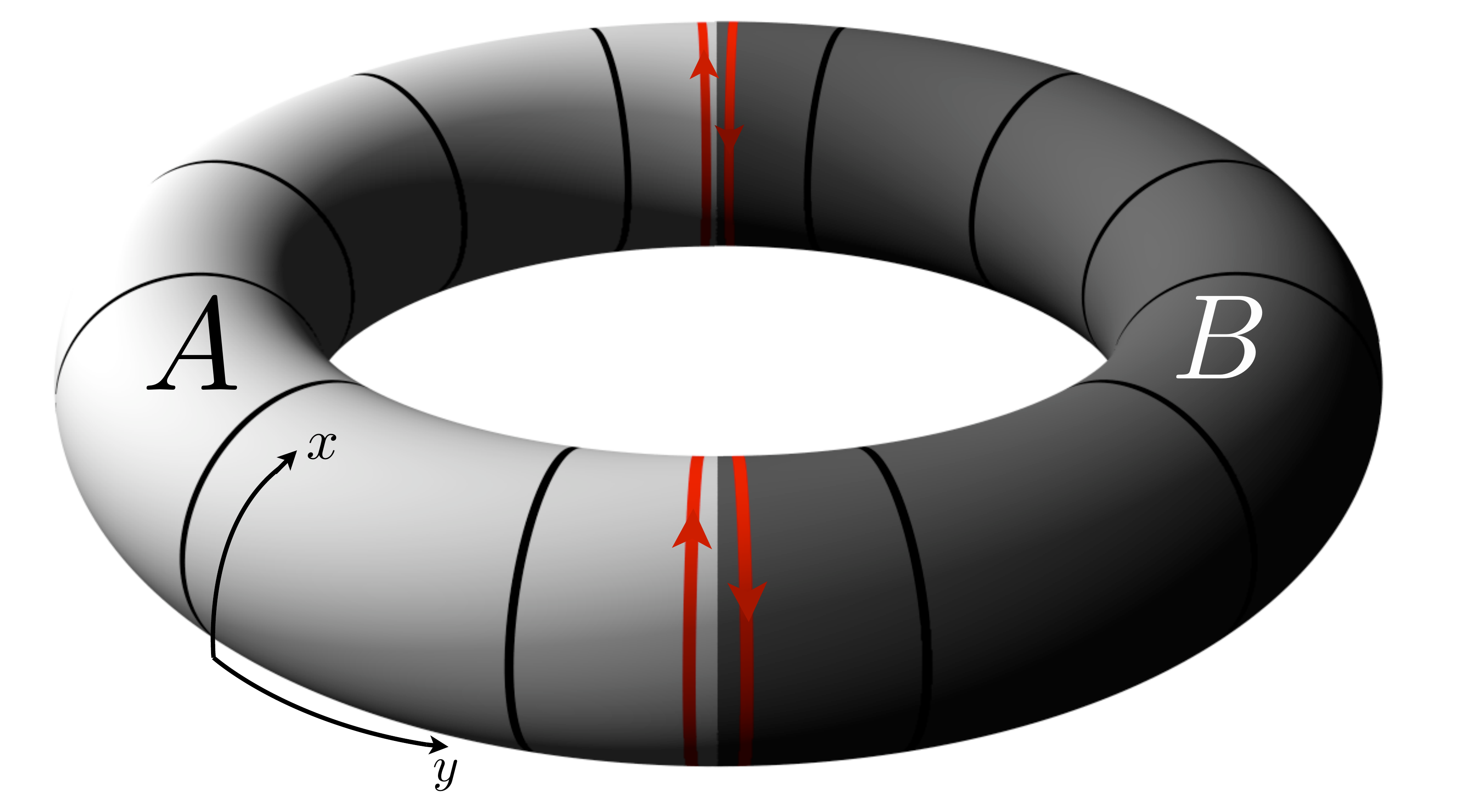}}
\caption{ (Color online) Torus setup for block entanglement computations.  The
lowest Landau level is spanned by orbitals which in Landau gauge are centered
along the circles shown.  The arrows indicate the chiralities of the virtual
`edges' created by the block partitioning.
\label{fig:torus}}
\end{figure}
%%%%%%%%%%%%%%%%%%%%%%%%%%

Very recently, ES studies have been used \cite{LiHaldane_PRL08,
  RegnaultBernevigHaldane_PRL09} to probe edge modes of FQH states.  The
entanglement between two partitions of an edgeless wavefunction may seem at
first sight unrelated to edge physics.  However, studies of ES in
non-interacting systems \cite{Peschel, Fidkowski} have found that the
entanglement spectrum is also the spectrum of an effective ``entanglement
Hamiltonian'' confined to the $A$ region of space, which is not identical but
similar to the original physical Hamiltonian. If this similarity holds for
interacting systems, 
%the low-lying structure of the ES would be similar to the
%low-energy spectrum of a system confined to the region $A$.  Since the region
%$A$ does have a boundary, 
the low-lying ES would then show an edge
structure, even though the total system has no edge.
Refs.~\cite{LiHaldane_PRL08, RegnaultBernevigHaldane_PRL09} analyzed the ES of
FQH states on the sphere with hemispheric partitioning.  The Virasoro
multiplet structure of the conformal field theory (CFT) describing the edge
appear in the low-lying part of the ES.

%%%%%%%

In this Letter we present and analyze the entanglement spectrum of $\nu=1/3$
Laughlin states on the torus.  This choice of geometry gives us access to new
physics and new analysis tools, compared to the spherical case.
The natural partitions of the torus are cylinder-like segments with two
disjoint edges.  The ES thus contains the physics of a combination of two
separate conformal edges.  We show that this leads to `towers' in the ES
spectrum, when plotted against appropriate quantum numbers.  
Even in cases where the two edges have different spectra, the two spectra
combine to form towers. 
%
%% To the best of our knowledge, this is the only explicit example of formation
%% of conformal towers from two separated chiral edges.
%
The two-edge picture provides significant predictive power, as the
assignment of only a few edge mode energies enables us to construct the
remaining ES.

The torus geometry also allows us to adiabatically connect to the ``thin
torus'' limit, which is exactly solvable \cite{bk,seidel} and has as ground
states the Tao-Thouless crystalline states \cite{tt}.  Many features of the ES
can be understood starting from these simple states, such as the positions of
towers and relationships between their energetics.
The CFT tower structure persists even very close to the thin-torus limit,
which by itself is an uncorrelated product state.

%%%%%%%%%%%%%%%%%%%%%%%%%%
\begin{figure*}[h!t]
\centerline{
\includegraphics[width=0.79\linewidth]{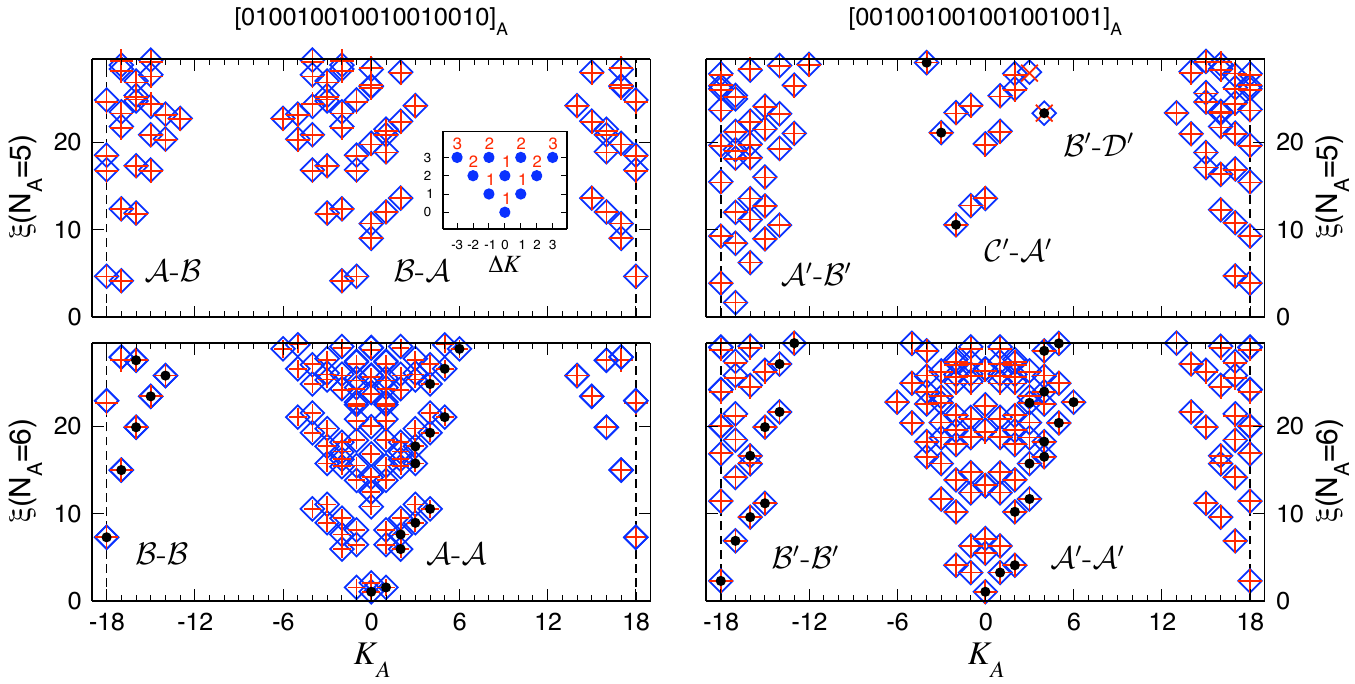}
}
\caption{ (Color online) Entanglement spectrum for $N_s=36$ and $L_1=10$. Left
  panels show the symmetric cut and right panels show one of the asymmetric
  cuts.  The origin of $K_A$ is chosen to match the Tao-Thouless state.  The
  blue squares represent numerically obtained data.  The assigned edge modes
  are labeled by black dots while the combinations of those edges are marked
  by red crosses.  The script letters are microscopic identifiers for the two
  edges combining to form each tower (see text). The striking correspondence
  of the red crosses with numerical data shows that our algorithm based on the
  two-edge picture allows the reconstruction of the entire entanglement
  spectrum using only the positions of the black dots.  The inset shows a CFT
  tower formed by two ideal chiral edges, the states labeled with their
  degeneracies.
\label{fig:maindata}}
\end{figure*}
%%%%%%%%%%%%%%%%%%%%%%%%%%

\paragraph*{Geometry and partitioning ---}

We study an $N$-electron system on a torus with periods $L_1, L_2$ in the $x$-
and $y$-directions, satisfying $L_1L_2=2\pi N_s$ (in units of the magnetic
length).  Here $N_s=N/\nu$ is the number of magnetic flux quanta.
In the Landau gauge, $\mathbf{A}=By\mathbf{\hat{x}}$, a
basis of single particle states in the lowest Landau level can be taken as
$\psi_{j}=\pi^{-1/4}L_1^{-1/2}\sum_{m} e^{i(\frac{2\pi}{L_1}j+mL_2)x}
e^{-(y+\frac{2\pi}{L_1}j+mL_2)^2/2}$ with $j=0,1,...,N_s-1$. The states
$\psi_j$ are 
centered along the lines $y=-2\pi j/L_1$ (Fig.~\ref{fig:torus}).  Thus the
$y$-position is given by the $x$-momentum $j$.

Any translation-invariant two-body interaction Hamiltonian, acting in the
lowest Landau level, can be written as
\begin{equation}
\label{ham}
H =\sum_n \sum_{k > |m|} V_{km}c^\dagger_{n+m}c^\dagger_{n+k}c_{n+m+k}c_n \ \ ,
\end{equation}
where $c^\dagger_m$ is creates an electron in the state $\psi_m$ and $V_{km}$
is the amplitude for two particles to hop symmetrically from separation $k+m$
to $k-m$.  Hence, the problem of interacting electrons on a Landau level maps
onto a one-dimensional, center-of-mass conserving, lattice model with lattice
constant $2\pi/ L_1$.  The Laughlin states at $\nu=1/3$ are, for all $L_1$, the unique zero energy grounds states of
the pseudo potential interaction
$V^{(1)}_{km}=(k^2-m^2)e^{-2\pi^2(k^2+m^2)/L_1^2}$ \cite{bk}.
% , which gives the lattice version of Haldane's pseudopotential
% $V^{(1)}(\mathbf r)$ \cite{pseudopot}.
For generic interactions, \emph{e.g.}, the Coulomb interaction, the matrix
elements have a more complicated $L_1$ dependence. 
Our ES data are extracted from ground states of \eqref{ham}, obtained using
the Lanczos algorithm for numerical diagonalization.

We bipartition the system into blocks $A$ and $B$ which consist of $l_A$
consecutive orbitals and the remaining $N_s-l_A$ orbitals, respectively
(Fig.~\ref{fig:torus}).  Since the orbitals are localized, this is a
reasonable approximation to spatial partitioning, as on the sphere
\cite{HaqueZozulyaSchoutens,LiHaldane_PRL08,ZozulyaHaqueRegnault,RegnaultBernevigHaldane_PRL09}.
In this Letter we focus on half-partitioning, $l_A=N_s/2$ and organize the ES in sectors labeled by the particle
number, $N_A$, and the total $x$-momentum (along the block boundary), $K_A \ mod \ N_s$, in the $A$ block.  
%The latter
%(momentum along block boundary direction) is periodic with period $N_s$.

The $\nu=1/3$ Laughlin state is three-fold degenerate on the torus.
The degenerate states are related by translation, and correspond to three
different thin torus configurations: 
%for four particles ($N_s=12$) they are
%
\begin{gather*}
0\;1\;0\,\big{|}\,0\;1\;0\;0\;1\;0\,\big{|}\,0\;1\;0\;  \\
1\;0\;0\,\big{|}\,1\;0\;0\;1\;0\;0\,\big{|}\,1\;0\;0\;  \\
0\;0\;1\,\big{|}\,0\;0\;1\;0\;0\;1\,\big{|}\,0\;0\;1\;
\end{gather*}
These states are, for generic (including Coulomb and pseudo potential) interactions, adiabatically connected to the bulk ground states without gap closing for any $L_1$ \cite{bk,seidel}.
%
%% \[
%% 010\big{|}010010\big{|}010 \, ; \quad 100\big{|}100100\big{|}100 \, ; \quad 001\big{|}001001\big{|}001 \, .
%% \]
%
If the A partition is taken to be the six middle orbitals, the block
boundaries are different for the three cases: 0-0 cuts in the first case
(`symmetric cut'), a 1-0 and a 0-1 cut in the other two (`asymmetric cuts').
Thus the degenerate ground state wavefunctions have different ES's (which can equivalently be obtained by using a single
wavefunction, and placing the boundaries at three inequivalent positions).

\paragraph*{Tower structure and CFT identification ---}

Numerical ES are shown in Fig.~\ref{fig:maindata} for a 12-particle Laughlin
state ($N_s=36$).  
A prominent feature is that the ES consists of `towers'.
Most of the towers are symmetric, while some are skewed.
We interpret each tower as a combination of chiral modes of two edges (two
block boundaries).  An \textit{ad hoc} assignment of a small number of
(Virasoro) energies provides the necessary input for constructing each tower.

The number of independent modes of a chiral $U(1)$ CFT at momentum $k$ is
given by the partition function
$p(k)=
1,1,2,3,5,7,11,...$ for $k=$ 0,1,2,....
When two \emph{linearly} dispersing chiral modes combine, one expects an ideal
tower of states like the one shown in Fig.~\ref{fig:maindata}~(inset).  The
diagonal sequences along the left and right sides are from the individual
edges, and so have degeneracies $p(|{\Delta{K_A}}|)$ at momentum shifted by
$\Delta{K_A}$ from the tower center.  A right-moving mode at energy $E_1$ and
momentum $k_1$ and a left-moving mode at $E_2$ and momentum $-k_2$ will
combine to give a mode at energy $E_1+E_2$ and momentum $k_1-k_2$.  The rest
of the tower is obtained from the states of the two edges through such
combinations.
Here energies are measured with respect to the vacuum state at ${\Delta}K_A=0$.

The observed towers in the numerical ES can be explained by postulating the
individual edge spectra to have split degeneracies while preserving the
Virasoro counting.  Two such modified edge spectra can be combined to
construct towers which are less degenerate than the ideal case of
Fig.~\ref{fig:maindata}~(inset).
Following this scheme, for the numerical ES towers (Fig.~\ref{fig:maindata})
we assigned appropriate levels on the right and left of each tower to
single-edge spectra.  Several edges in the $N_A{\neq}N/2$ sectors are
identical to edges in the $N_A=N/2$ sector, and thus do not need to be
independently assigned.  As a result, the energies of a remarkably small
number of single-edge states (black dots in Fig.~\ref{fig:maindata}) are
sufficient to generate the entire ES.  This is a key result of the present
work.

\begin{figure}[t!]
\centerline{
\includegraphics[width=0.89\linewidth]{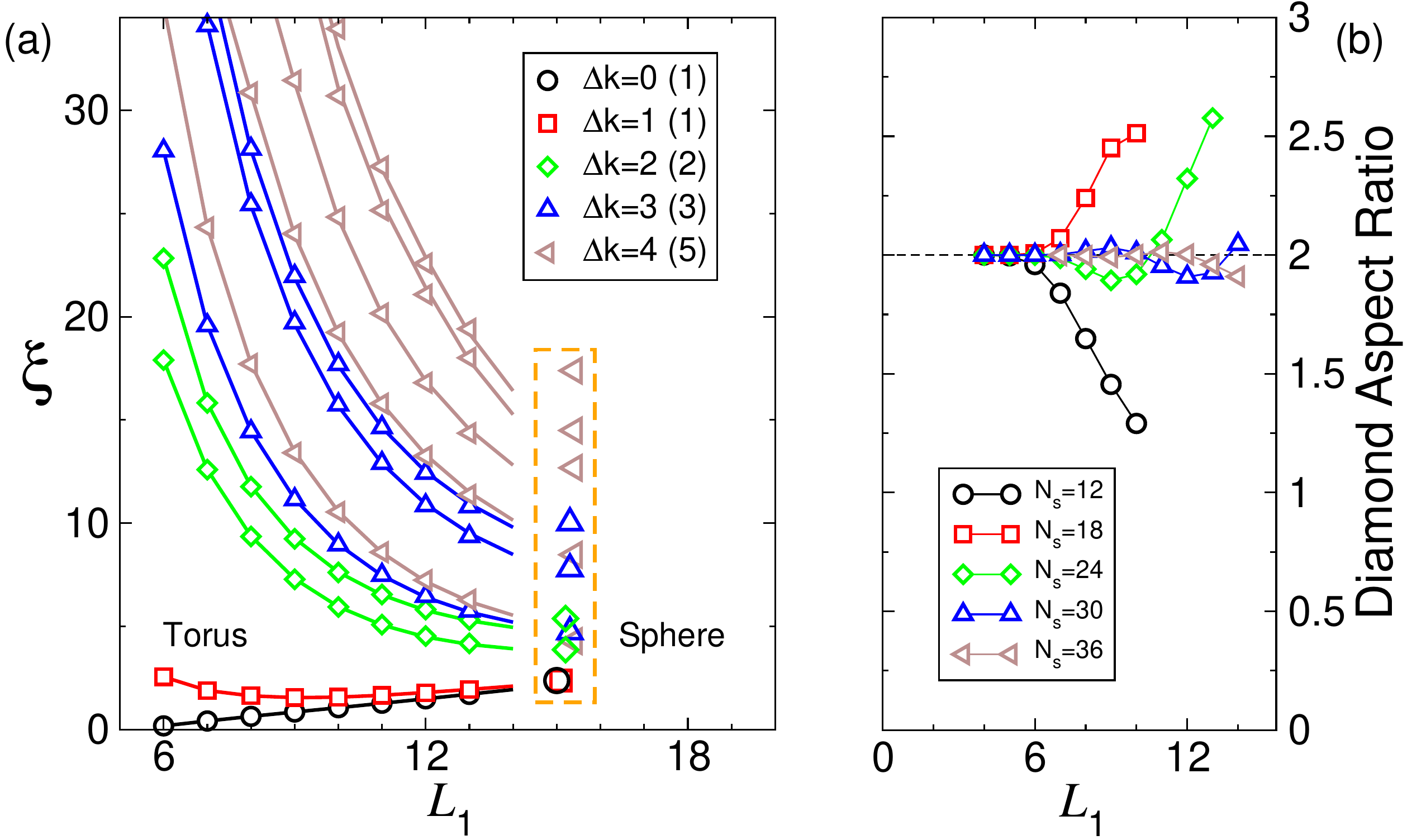}
}
\caption{ (Color online) (a) The chiral edge levels (type $\mathcal{A}$)
 identified from ES, as a function of torus thickness $L_1$.  The rectangle
 contains the single-edge ES levels in spherical geometry
 \cite{ZozulyaHaqueRegnault}, here scaled and shifted for best comparison with
 the torus results around $L_1\sim14$.
(b) `Aspect ratio' of the diamond formed by the four lowest ES levels of the
 $\mathcal{A}$-$\mathcal{A}$ tower.  The value 2 means a perfect diamond
 shape.
\label{fig:edgeLevels_DiamondAR}}
\end{figure}

The assigned single-edge levels have robust relative positions for varying
torus thickness (Fig.~\ref{fig:edgeLevels_DiamondAR}a).  The relative
positions correspond well to the single-edge levels extracted from ES on a
sphere \cite{ZozulyaHaqueRegnault}, as shown in
Fig.~\ref{fig:edgeLevels_DiamondAR}a.

%%%%%%%%%%%%%%%%%
%%%%%%%%%%%%%%%%%

\paragraph*{Microscopic `thin-torus' analysis ---}

The adiabatic connection to the thin-torus ($L_1\rightarrow 0$) limit enables
us to understand features of the ES by perturbing away from this solvable
limit.  In particular, the location and energetics of towers can be understood
from such microscopic considerations.

We first consider the symmetric cut and explain the towers in the left panels
of Fig. \ref{fig:maindata}. The very lowest ES level is found at $K_A=0$ in
the $N_A=6$ sector (left bottom panel), and corresponds to the parent
(thin-torus) configuration
\begin{equation}\label{TTroot} 
010010010 \big{|}\underbrace{010010010010010010}_{A}{\big |}010010010 \ \ .\nonumber
\end{equation} 
At $L_1=0$ this amounts to the only  entanglement level, and for all
finite $L_1$ its dressed counterpart remains at the very bottom of the main
tower. The remaining states of this tower are all generated from this by
$V_{km}$ processes (Eq.~\ref{ham}) which conserve $N_A$.  In particular, the
leading levels are generated by processes with small $k$ and $m$. For example,
a $V_{21}$ process at the right edge, $0010|0100 \rightarrow 0001|1000$; gives
the root configuration for the lowest entanglement level at $\Delta{K_A}=1$
and at the left edge this gives the lowest $\Delta{K_A}=-1$ level.  The
energetics of the tower is determined by the microscopic structure of its
edges.  We refer to this particular environment, $10010|01001$, and the
corresponding edge energetics as $\mathcal{A}$.

Some processes do not conserve $N_A$. For the symmetric cut the leading such
process is $V_{42}$ which can change $N_A$ by $\pm1$: $0010|01001 \rightarrow
0000|1110$.  We call such an edge environment $\mathcal{B}$, which combines
with $\mathcal{A}$ type edges to form the observed $\mathcal{A}$-$\mathcal{B}$
and $\mathcal{B}$-$\mathcal{A}$ towers in the $N_A=5$ sector.  Mirror images
of these exist in the $N_A=7$ sector.  By creating two $\mathcal{B}$ edges one
finds that there are two ways of obtaining the $\mathcal{B}$-$\mathcal{B}$
tower in the $N_A=6$ sector, each with momentum shift $\pm l_A$ ($= \pm N_s/2$)
compared to the main tower. The extra two-fold degeneracy is seen in our data.
The large momentum transfer is because a particle leaves block $A$ at one cut,
and one enters $A$ at the opposite cut. We also predict and observe a
nondegenerate $\mathcal{B}$-$\mathcal{B}$ tower in the $N_A=4$ sector (not
shown).

We now turn to the asymmetric cut (right panels of Fig.~\ref{fig:maindata}).
Again, the thin torus ground state
corresponds to the lowest level in the main tower (at $K_A=0$ in the $N_A=6$
sector). The two edges, $0100|1001$ and $1001|0010$, are both denoted
$\mathcal{A}'$ as they are each others mirror images and hence have equivalent
energetics. In this case already the leading hopping term, $V_{21}$, changes
$N_A$ leading to the $\mathcal{B}'$ edge, $0100|1001 \rightarrow 0011|0001$.
Another edge $\mathcal{C}'$ is generated by the $V_{54}$ process,
$1001001|001001 \rightarrow 0001111|000001$.  A feature of the ES for the
asymmetric cut is that the skewed towers within a given $N_A$ sector do not in
general have mirror image in that sector. Instead, the mirror images show up
in the $(N-N_A)$ sector.  Note that the energy of the lowest $\Delta{K_A}=0$
state for each tower (tower vacuum energy) is also fixed by the two edges.

Pursuing the microscopic analysis, one can find many non-trivial relations
between the energetics at different towers, cuts and sectors; we have only
outlined the basics.  It is also possible to derive more quantitative features
of the ES as a function of $L_1$, through perturbative calculations starting
at the thin-torus limit.  Details will be explored elsewhere.

\paragraph*{Circumference and size dependence ---}

\begin{figure}[t!]
\centerline{\includegraphics*[width=0.89\linewidth]{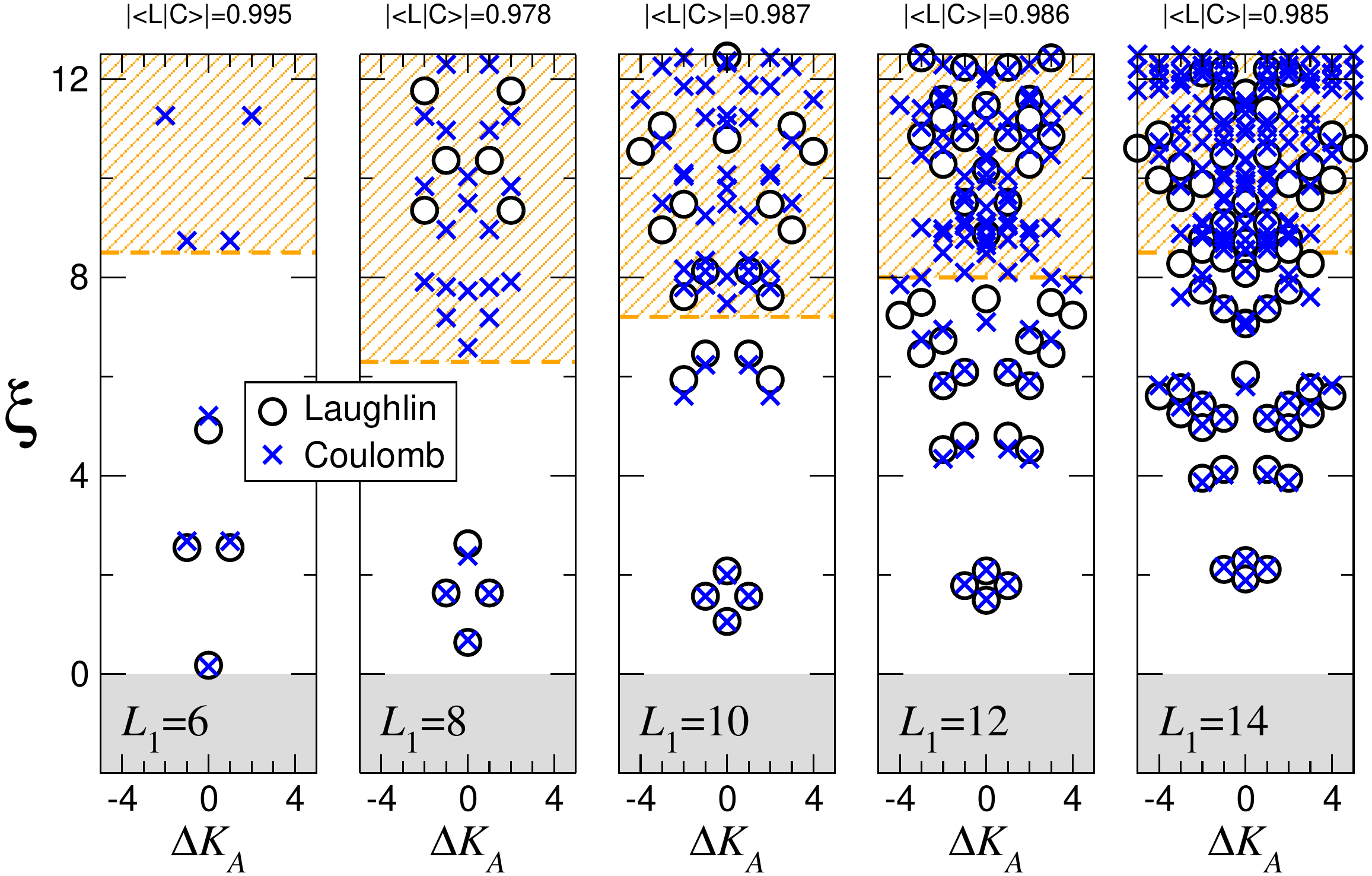}}
\caption{ (Color online) Comparison of the entanglement spectrum, and overlaps, between the
Laughlin wave function and the Coulomb ground state for $L_1=6,\ldots ,14$ at
fixed $N_s=36$. Only the central part of the most prominent tower for the
symmetric cut is displayed. We observe that the entanglement spectra of the
two wave functions become very similar for sufficiently large $L_1$. 
The appearance of "generic levels" beyond the two-edge CFT picture in the Coulomb state is indicated by the shaded regions,
leading to a tentative notion of "entanglement gap" \cite{LiHaldane_PRL08}.
\label{fig:espec_onethird_laughlin_coulomb_L1}}
\end{figure}

We illustrate the roles of $N_s$ and $L_1$ by focusing on the four lowest
levels of the $\mathcal{A}$-$\mathcal{A}$ tower, which form a diamond shape
(\emph{e.g.}, Fig.~\ref{fig:espec_onethird_laughlin_coulomb_L1}).  The ratio
of the energy of the second $\Delta{K_A}=0$ level to the first $\Delta{K_A}=1$
level, each measured from the lowest $\Delta{K_A}=0$ level, is an `aspect
ratio' for the diamond shape, and is plotted in
Fig.~\ref{fig:edgeLevels_DiamondAR}b.  Both the two-edge CFT picture and
perturbation from the thin-torus limit predict aspect ratio $=2$.  This is seen
to hold from the thin-torus limit up to some threshold value of $L_1$, which increases as $N_s$ is
increased.  
This situation is generically true for all CFT features: although the Laughlin
state converges to CFT behavior at any $L_1$, at larger circumferences more
particles are required for finite-size convergence.

For large $L_1$ and finite $N_s$, the edges are close and therefore interact,
leading to complicated effects such as the aspect ratio deviations seen in
Fig.~\ref{fig:edgeLevels_DiamondAR}b.

\paragraph{Coulomb ground states ---}
The ES for ground states of the Coulomb Hamiltonian have more complicated
$L_1$ dependence (Fig.~\ref{fig:espec_onethird_laughlin_coulomb_L1}).  At
smaller $L_1$ only the few lowest levels resemble the Laughlin ES; for
$L_1\lesssim8$ the Coulomb ES cannot be generated using the two-edge procedure
beyond the diamond structure. This does however not contradict the larger overlap 
between Laughlin and Coulomb states known at $L_1\rightarrow 0$ \cite{Yang, bk}, 
because the higher ES levels are pushed upwards at small $L_1$ ({\it cf} also 
Fig.~\ref{fig:espec_onethird_laughlin_coulomb_L1}).  The ES thus exposes correlations
much more subtle than is visible in overlap considerations.
As $L_1$ is increased, more and more Coulomb ES levels match the Laughlin ES, and 
the emergent CFT tower structure can be seen.

%%%%%%%%%%%%%%%%%
%%%%%%%%%%%%%%%%%

\paragraph*{Discussion ---}

This work presents entanglement spectra calculated through numerical
diagonalization, for the Laughlin state at $\nu=1/3$ on a torus.  We presented
two radically different physical ways of understanding the ES structure.
The first interpretation is based on a combination of two chiral CFT edges.
Each of these are individually similar to the edge spectrum previously
extracted from ES studies on the sphere \cite{ZozulyaHaqueRegnault}.  This
interpretation
is powerful as it reproduces the entire ES through the assignment of a
few levels.
Our second approach uses the adiabatic connection to the thin-torus limit, and
the remarkable fact that the two-edge CFT structure is preserved even close to
the thin-torus limit.  Perturbative analysis based on the simple thin-torus
states yields the locations and shapes of the towers, and many other
quantitative predictions.

Accessing edge modes in explicit numerical calculations remains a highly
desired but difficult task, due to edge reconstruction and other difficulties
\cite{edgeNumerics_XinWan_Jain}.  Our study of edge combinations through
entanglement calculations provides an alternative track to gaining insight
into this issue.

Our work opens up several important research directions, of which we list a
few.  We expect our results to have interesting generalizations to more
intricate FQH states, such as the non-abelian states. 

Our data at large $L_1$ deviates from the independent-edges picture because
the edges are close.  The present setup thus provides the intriguing
possibility of studying the interaction and interferences between two
spatially separated edges. \emph{e.g.}, through exploring large $L_1$ features
as in Fig.~\ref{fig:edgeLevels_DiamondAR}b.  

The CFT edge interpretation relies on the idea that the `entanglement
Hamiltonian'
is similar to the physical Hamiltonian.  This notion is plausible but entirely
unexplored for FQH states.  There is thus a clear need for constructing and
understanding entanglement Hamiltonians.  It is also possible that a more
detailed study of the CFT towers in the ES could  yield Luttinger liquid features
such as the compactification radius and more generally the scaling dimensions.

\acknowledgements
We acknowledge ZIH TU Dresden and MPG RZ Garching for allocation of computing time.


\begin{thebibliography}{99}


\bibitem{AmicoFazioOsterlohVedral_RMP08} L. Amico  {\it et\ al.}, %, R. Fazio, A. Osterloh, and V. Vedral,
  Rev.\ Mod.\ Phys.\ {\bf 80}, 517 (2008).

\bibitem{kitaev06} A.~Kitaev and J.~Preskill, Phys.\ Rev.\ Lett.\ {\bf 96}, 110404-1 (2006).
\bibitem{levin06} M.~Levin and X.~G.~Wen, Phys.\ Rev.\ Lett.\ {\bf 96}, 110405 (2006).

\bibitem{LiHaldane_PRL08}  H.~Li and F.~D.~M.~Haldane, Phys.\ Rev.\ Lett.\
 {\bf 101}, 010504 (2008). 

\bibitem{topol-quantum-computing} 
M.~Freedman, M.~Larsen, and Z.~Wang, Commun.\ Math.~Phys.\ {\bf 227}, 605
(2002).  \, 
%S.~Das Sarma, M.~Freedman, and C.~Nayak, \emph{Physics Today}, July 2006,
%page 32. \,
%
% A.~Yu. Kitaev, Ann. Phys. {\bf 303}, 2 (2003). \,
%
C.~Nayak \emph{et\ al.},
%  C.~Nayak, S.~H.~Simon, A.~Stern, M.~Freedman, S.~Das Sarma, 
Rev.\ Mod.\ Phys.\ {\bf 80}, 1083 (2008). 


\bibitem{Wen_chiraledges} X.~G.~Wen, Phys.\ Rev.\ B {\bf 41}, 12838 (1990);
{\bf 43}, 11025 (1991); {\bf 44}, 5708 (1991).


\bibitem{edgeNumerics_XinWan_Jain} X.~Wan, K.~Yang, and E.~H.~Rezayi, 
Phys.\ Rev.\ Lett.\ {\bf 88}, 056802 (2002).  \,
%
X.~Wan, E.~H.~Rezayi and K.~Yang, 
Phys.\ Rev.\ B {\bf 68}, 125307 (2003). \,
%
S.~Jolad and J.~K.~Jain, Phys.\ Rev.\ Lett.\ {\bf 102}, 116801 (2009).


\bibitem{RegnaultBernevigHaldane_PRL09} 
N.~Regnault, B.~A.~Bernevig and F.~D.~M.~Haldane,  Phys.\ Rev.\ Lett.\
{\bf 103}, 016801 (2009).

\bibitem{Peschel} M.-C. Chung, and I. Peschel, Phys. Rev. B {\bf 64}, 064412
 (2001). \,
%
I.~Peschel and V.~Eisler, J. Phys. A: Math. Theor. {\bf 42}, 504003 (2009).

\bibitem{Fidkowski} L.~Fidkowski,  arXiv:0909.2654v2.


\bibitem{bk}
	 E.~J.~Bergholtz and A.~Karlhede,  Phys.\ Rev.\ Lett.\ {\bf 94}, 026802 (2005);
	 J.~Stat.\ Mech.\ L04001 (2006);
	 Phys.\ Rev.\ B {\bf 77}, 155308 (2008); E.~J.~Bergholtz {\it et\ al.}, %  , T.~H.~Hansson, M.~Hermanns and A.~Karlhede,
	 Phys.\ Rev.\ Lett.\ {\bf 99}, 256803 (2007).

\bibitem{seidel}
% A.~Seidel, H.~Fu, D.~-H.~Lee, J.~M.~Leinaas, and J.~Moore,
A.~Seidel {\it et\ al.}, 
Phys.\ Rev.\ Lett.\ {\bf 95}, 266405 (2005).

\bibitem{tt}
	 R.~Tao and D.~J.~Thouless,
	 Phys.\ Rev.\ B {\bf 28}, 1142 (1983).

% \bibitem{pseudopot} F. D. M.  Haldane, Phys. Rev. Lett. {\bf 51}, 605 (1983). 

\bibitem{HaqueZozulyaSchoutens} M.~Haque, O.~Zozulya and K.~Schoutens, Phys.\ Rev.\ Lett.\
 {\bf 98}, 060401 (2007).  \, 
%
O.~S.~Zozulya, {\it et\ al.},  % M.~Haque, K.~Schoutens and E.~H.~Rezayi,
Phys.\ Rev.\ B {\bf 76,} 125310 (2007). \,
%
\bibitem{ZozulyaHaqueRegnault}  O.~Zozulya, M.~Haque and N.~Regnault, Phys.\ Rev.\ B {\bf 79} 045409 (2009).

\bibitem{Yang}K.~Yang, F.~D.~M.~Haldane and E.~H.~Rezayi,
	 Phys.\ Rev.\ B {\bf 64}, 081301(R) (2001).

\end{thebibliography}
\end{document}